\documentclass[preprint,preprintnumbers,amsmath,amssymb]{revtex4}

\usepackage{graphicx}
\usepackage{dcolumn}
\usepackage{bm}
\usepackage[latin1]{inputenc}
\usepackage{upgreek}

\begin{document}

\title{Spin-wave interference in three-dimensional rolled-up ferromagnetic microtubes}

\author{Felix Balhorn}
\author{Sebastian Mansfeld}
\author{Andreas Krohn}
\author{Jesco Topp}
\author{Wolfgang Hansen}
\author{Detlef Heitmann}
\author{Stefan Mendach}
\altaffiliation[corresponding author, email:
]{smendach@physnet.uni-hamburg.de}

\affiliation{%
Institut f\"{u}r Angewandte Physik und Zentrum f\"{u}r Mikrostrukturforschung, Universit\"{a}t Hamburg, Jungiusstrasse 11, D-20355 Hamburg, Germany}%

\date{\today}

\begin{abstract}
We have investigated spin-wave excitations in rolled-up Permalloy
microtubes using microwave absorption spectroscopy. We find a series
of quantized azimuthal modes which arise from the constructive
interference of Damon-Eshbach type spin waves propagating around the
circumference of the microtubes, forming a spin-wave resonator. The
mode spectrum can be tailored by the tube's radius and number of
rolled-up layers.
\end{abstract}

\maketitle
\newpage

Spin-wave excitations in thin films and patterned ferromagnetic
structures are of both fundamental and practical scientific
interest. On the micrometer scale geometric boundaries induce
quantization conditions on spin-wave excitations. This effect has
been shown for flat laterally patterned geometries such as wires and
rectangular elements
\cite{PhysRevLett.88.047204,PhysRevB.69.134401,Demidov2008} as well
as for micrometer sized rings and discs
\cite{jorzick:3859, PhysRevLett.95.167201,giesen:014431,podbielski:167207,Back2006}.\\
\indent In this letter we report on the fabrication and experimental
investigation of novel three-dimensional ferromagnetic
microstructures consisting of rolled-up Permalloy/semiconductor
($\textrm{Ni}_{80}\textrm{Fe}_{20}/\textrm{GaAs}/\textrm{In}_{20}\textrm{Ga}_{80}\textrm{As}$)
layers, in the following termed \underline{r}olled-\underline{u}p
\underline{P}ermalloy micro\underline{t}ube (RUPT). Using microwave
absorption spectroscopy on homogenously magnetized RUPTs we observe
a series of sharp modes which arise from the constructive
interference of Damon-Eshbach type~\cite{Damon} spin waves traveling
along the circumference of the RUPT. These interference conditions
are quite universal. They resemble the acoustic whispering-gallery
modes in the St. Pauls Cathedral, originally explored by Lord
Rayleigh \cite{Rayleigh1912}, optical resonators based on
semiconductor microtubes \cite{strelow:127403, Mendach08} and
microdiscs
\cite{mccall:289,PhysRevLett.95.067401} or dielectric microspheres~\cite{Vernooy1998, nature001}.\\
\indent The measurements discussed in this work were performed on two RUPTs with different radii
$r$, winding numbers $N$ (giving the number of revolutions of the rolled-up material) and layer
composition (RUPT A and RUPT B, parameters given below). They were prepared from a strained
Permalloy/semiconductor multilayer, utilizing the self-rolling effect \cite{Cho2006} pioneered
by Prinz et al. \cite{p2} and Schmidt et al. \cite{schmidt2001}.\\
\indent For RUPT A (RUPT B) a $t=20$~nm thick Permalloy layer was
thermally evaporated on a molecular beam epitaxy grown
heterostructure made of $15$~nm ($10$~nm) $\textrm{Ga}\textrm{As}$,
$15$~nm $\textrm{In}_{20}\textrm{Ga}_{80}\textrm{As}$ and $40$~nm
$\textrm{Al}\textrm{As}$ on a $\textrm{Ga}\textrm{As}$ substrate
(see Fig. 1(a)). The lattice constant of $\textrm{In}\textrm{As}$ is
significantly larger than for $\textrm{Ga}\textrm{As}$ and allows a
pseudomorphically strained growth. By selectively etching the AlAs
sacrificial layer the
$\textrm{Ni}_{80}\textrm{Fe}_{20}/\textrm{GaAs}/\textrm{In}_{20}\textrm{Ga}_{80}\textrm{As}$
 layer system is released from the substrate and minimizes its strain energy by rolling up into a tube (see Fig. 1(b) and (c)).
The diameter $d$ of the RUPT is determined by the composition and
thickness of the
$\textrm{Ni}_{80}\textrm{Fe}_{20}/\textrm{GaAs}/\textrm{In}_{20}\textrm{Ga}_{80}\textrm{As}$
layer system \cite{Grundmann2003}. The length $l$ and winding number
$N$ of the RUPT can be precisely controlled by photolithography,
i.e. by defining the lateral dimensions of a strained mesa which is
rolled-up in the final selective etching step from a well-defined
starting edge \cite{Schumacher2005}. After the measurements were
performed on RUPT \nolinebreak A, a window was prepared into its
surface by focused ion beams to enable an analysis of its inner
structure  (see inset in Fig. 1(c)). The RUPT is tightly rolled up,
spiral in shape and formed from $3.5$ windings of the
$\textrm{Ni}_{80}\textrm{Fe}_{20}/\textrm{GaAs}/\textrm{In}_{20}\textrm{Ga}_{80}\textrm{As}$
multilayer. Together with each RUPT an unpatterned Permalloy film
from the same wafer was prepared, using the same thermal evaporation
step to allow the determination of reference parameters. The film
for RUPT A (RUPT B) has a saturation magnetization
$\textrm{M}_{\textrm{S}} = 980$~mT ($\textrm{M}_{\textrm{S}} =
1080$~mT) and a Gilbert damping constant $\alpha=0.008$
($\alpha=0.008$), as determined
by microwave absorption measurements.\\
\indent We investigated the spin-wave spectrum using high-resolution
microwave absorption spectroscopy. For this purpose the RUPT was
removed from the GaAs substrate and placed on the $2.4~\upmu$m wide
signal line of a coplanar waveguide (CPW) using a setup with piezo
controlled manipulation needles \cite{Mendach2009}. The CPW was
defined by optical lithography on a GaAs wafer and consists of a
$170$~nm thick trilayer of Cr/Ag/Au. A static external magnetic
field $\vec{H}(x,y)$ was applied in the plane of the waveguide along
the tube's axis. With a vector network analyzer (VNA) we measured
the microwave transmission through the CPW in dependence of the
microwave frequency~$f$. The high-frequency magnetic field of the
CPW $h(y,z)$ pointed perpendicularly to the axis of the RUPT. The
excitation of a spin wave is indicated by a reduced transmission of
the waveguide due to the absorption of power at the corresponding
excitation frequency~$f$. Each measurement was normalized to a
reference measurement, taken with
$\mu_{\text{0}}{H}_{\text{ref}}=90$~mT applied perpendicularly to
the axis of the RUPT. To assure a well-defined magnetization
configuration within the RUPT, it was magnetized along its axis with
$\mu_{\text{0}}{H}_{\text{1}}=50$~mT prior to every frequency sweep,
ramped down to zero and then ramped up to the actual field. For
a detailed description of the technique used in this experiment, see Ref. \cite{giesen:014431}.\\
\indent Let us first discuss measurements performed on RUPT
\nolinebreak A with diameter $d_{\textrm{A}} = 3.5~\upmu
\textrm{m}$, length $l_{\textrm{A}}= 60~\upmu \textrm{m}$ and
rolling number $N_{\textrm{A}} = 3.5$.  Figure 2(a) shows the
excitation spectrum of RUPT A at an external magnetic field $H = 0$.
Four distinct resonances are observed at
$f_{\text{\text{A,0}}}=3.5$~GHz, $f_{\text{\text{A,1}}}=4.5$~GHz,
$f_{\text{\text{A,2}}}=5.3$~GHz, and $f_{\text{A,3}}=5.8$~GHz. With
increasing external magnetic field the spacing between all resonance
peaks reduces until they overlap. To determine the exact
eigenfrequency of each spin wave each peak was fitted assuming a
Lorentzian shaped curve. Figure 2(b) depicts a plot of the resonance
frequency of the three larger resonance peaks for different external
magnetic fields $H$. Figure \nolinebreak 2(c) and (d) display
corresponding data for RUPT \nolinebreak B with diameter
$d_{\textrm{A}} = 2.8~\upmu \textrm{m}$, length $l_{\textrm{A}}=
100~\upmu \textrm{m}$ and winding number $N_{\text{B}} = 1.8$. This
RUPT shows four resonance peaks for $H=0$ at
$f_{\text{\text{B,0}}}=4.1$~GHz, $f_{\text{\text{B,1}}}=5.1$~GHz,
$f_{\text{\text{B,2}}}=6.1$~GHz, and $f_{\text{B,3}}=6.7$~GHz. RUPT
B absorbs about twice as much microwave power as RUPT A.\\
\indent In the following we show that the resonances in the spectra
of RUPT A and RUPT B are due to spin waves traveling around the
RUPTs perimeter. They form resonant modes $n=0,1,2,3,...$ if the
periodic boundary condition for a ring resonator
\begin{equation}\label{eqn_resonance}
n \lambda =  \pi \cdot d \Leftrightarrow k_{\phi}=2n/d, \quad n \, \epsilon \, \mathbb{N}_{0}
\end{equation}
is fulfilled. Here $\lambda$ denotes the wavelength of the spin
wave, $k_{\phi}$ the azimuthal wave vector and $d$ the diameter of
the RUPT. Condition (\ref{eqn_resonance}) is true if an integer
number of wavelengths fits into the circumference of the RUPT, so
that spin-waves interfere constructively. We prepared RUPT B on
purpose with smaller diameter and a winding number of only $1.8$ to
confirm our explanation. The fact that we observe four resonant
modes for RUPT B as well confirms that they originate from
interfering azimuthal spin waves and not from individual isolated
modes in single films or parts of the circumference. In particular
we observe that with decreasing diameter $d=2.8~\upmu \textrm{m}$
for RUPT B as compared to $d=3.5~\upmu \textrm{m}$ for RUPT A the
resonance frequency increases as expected from the spin-wave model
described below.\\
\indent At first it seems surprising that spin waves interfere on
the circumference, since the actual ferromagnetic film is of spiral
and not closed shape. The explanation is that the individual
overlapping films couple strongly through dipole-dipole interaction.
This coupling also automatically leads to a renormalization of the
spin-wave mode frequencies as compared to a single-layered film.
This can be modeled by a single-layered tube with an effective
thickness $t_{\textrm{eff}}$. Additionally, the curved shape of the
microtubes also leads to both static and dynamic demagnetization
effects \cite{Mendach2009}. This, the \emph{demagnetization},
\emph{coupling} and \emph{renormalization} and their incorporation
into a \emph{spin-wave model} will be addressed in detail in the
following.\\
\indent \emph{Spin-wave model}. --- To describe our data, we start
from a model introduced by Kalinikos and Slavin \cite{Kalinikos1986}
for a flat, thin film, modified by Guslienko et al. \nolinebreak
\cite{PhysRevB.68.024422}. Here, the resonance frequency is given by
\begin{align}\label{eqn_dispersion}
f({k}_{\phi},{H}_{\text{int}})=&\frac{\gamma \mu_{0}}{2 \pi} \sqrt{\left[{H}_{\text{int}}+\frac{2A}{M_{\text{s}}} {k}_{\phi}^{2} \right]} \cdot \\
& \sqrt{\left[{H}_{\text{int}}+\frac{2A}{M_{\text{s}}} {k}_{\phi}^{2} + M_{s} F({k}_{\phi}, {H}_{\text{int}}) \right]}. \nonumber
\end{align}
$H_{\textrm{int}}$ is the internal magnetic field, $A$ is the
exchange constant for Permalloy ($A=13 \cdot 10^{-12} \frac{J}{m}$),
$M_{\text{s}}$ is the saturation magnetization and
$F({k}_{\phi},{H_{\text{int}}})$ is the dipole-dipole interaction
matrix given by
\begin{eqnarray}
  F({k}_{\phi},H_{\text{int}}) = 1 + P({k}_{\phi}) (1-P({k}_{\phi})) \left(\frac{M_{\text{s}}}{{H}_{\text{int}}+\frac{2A}{M_{\text{s}}}{k}_{\phi}^{2}}\right)&
\end{eqnarray}
with
\begin{equation}\label{thickness}
P({k}_{\phi})= 1 - \frac{1-e^{-k_{\phi}t_{\textrm{eff}}}}{{k}_{\phi}t_{\textrm{eff}}}.
\end{equation}
Spin waves traveling around the perimeter of an axially magnetized
RUPT with an external magnetic field parallel to its axis exhibit
Damon-Eshbach character for all  field values. The dispersion of
these spin waves calculated with equation (\ref{eqn_dispersion}) is
shown in Fig. 3. From the quantized wave vector $k_{\phi}$ given by
equation (\ref{eqn_resonance}) we directly get the corresponding
resonance frequency. We see that indeed the spacing between the
modes decreases with mode number as observed in the experiment. We
used this model to fit our data. The coupling and frequency
renormalization were accounted for by an effective layer thickness
$t_{\textrm{eff}}$. As explained below, the demagnetization was
considered by adding an effective demagnetizing field to the
internal magnetic field, $H_{\textrm{int}} = H + H_{\textrm{dem}}$.
We find a nearly perfect agreement with the experimental data using
$t_{\textrm{eff}}=37$~nm and $H_{\textrm{dem}}=16$~mT for RUPT A and
$t_{\textrm{eff}}=30$~nm and $H_{\textrm{dem}}=20$~mT for RUPT B.
The fit is shown in Fig.
\nolinebreak 2(a) and (b).\\
\indent \emph{Demagnetization}. --- Due to the bent shape of the
RUPT any precession of the magnetization leads to magnetization
components pointing towards geometric boundaries. This causes
dynamic demagnetizing effects which form a two-dimensional potential
energy landscape for a precessing spin \cite{Mendach2009}. For this
reason, even for $H=0$ the RUPT has a resonance frequency $f\neq0$.
Thus, to explain our data we have to take into account an additional
effective demagnetizing field $H_{\textrm{dem}}$. This concept is
similar to the treatment of a thin ferromagnetic wire in the Kittel
formula \cite{Kittel1948}. For wires such an effective field can be
accurately retrieved from the hard-axis dispersion relation, i.e.
when external field and long axis of the wire are perpendicular to
each other. In this case, if the external field is large enough, it
forces the magnetization to align itself also perpendicularly. This
creates a static demagnetization field, which reduces the internal
field in the wire. At the external field for which the frequency of
the fundamental spin-wave mode is minimal, the external and the
demagnetization field compensate each other \cite{Bailleul2003}.
Using this value for $H_{\textrm{dem}}$ the resonance frequency and
dispersion of a thin wire can be modeled within sufficient accuracy.
We used an identical approach for the RUPTs. The experimentally
observed hard-axis dispersion indeed displays a pronounced minimum
at $\mu_{\text{0}}H=17$~mT for RUPT A and at $\mu_{\text{0}}H=21$~mT
for RUPT B (see Fig. 3(b) and (c)), which is quite close to the
values used above to fit our data.\\
\indent \emph{Dipole coupling}. --- The RUPT is of spiral shape so
that at first sight there is no periodic azimuthal path through
ferromagnetic material for the spin-waves to travel. Instead the
spin-waves traveling in individual layers of the RUPT couple via
dipole-dipole interaction. To support this approach, we performed
micromagnetic simulations using the OOMMF
framework~\cite{Donahue99}. The simulation of the bent shape of a
microtube is, however, extremely time consuming and beyond our
computer capacities due to the rectangular discretization of space
required for the simulations with OOMMF.  Instead, we simulated two
thin rectangular Permalloy stripes stacked vertically (separation
distance 35~nm). Only the lower stripe is excited directly by a
$2.6$~ps long pulse applied perpendicular to an external in-plane
field of $20$~mT. Our simulations show that the second, not directly
excited stripe follows the excitation of the lower stripe via
dipole-dipole coupling with almost the same
amplitude, indeed indicating strong dipole-dipole coupling.\\
\indent \emph{Renormalization}. --- Due to the coupling the spin
excitations can be treated as an one-layered spin-wave system,
described by an effective thickness. As discussed above, we find a
nearly perfect agreement using an effective thickness of
$t_{\textrm{eff}}=37$~nm (30~nm) for RUPT~A~(B). These values lie in
between the thickness of a single layer, $t=20$~nm, and the total
thickness of $80$~nm ($40$~nm) for RUPT~A~(B). Also in the
micromagnetic simulations we find that the resonance frequency of
the coupled two-layered stripe system is shifted to higher
frequencies with regard to the single-layered system, which can
indeed be described by an increased effective thickness.\\
\indent The fits in Fig. 2 agree extremely well with the
experimental data for RUPT~A, while for RUPT~B we find small
deviations. We attribute this to the fact that one part of RUPT~B
consists of two layers, while the other one consists of only one
layer. For this reason the difference between both parts is much
larger than the difference between the parts of RUPT~A with three
and four layers, which perhaps is the reason why the effective
thickness ansatz is less accurate here.\\
\indent In conclusion, we have fabricated and investigated novel
ferromagnetic microtube ring resonators. We observe quantized
spin-wave modes arising from the constructive interference on the
circumference of the tube. The mode spectrum can be tailored by the
tube's radius and number of windings. The confinement in these
microtube ring resonators is, in contrast to flat ferromagnetic
structures, not governed by geometrical edges and their complicated
magnetization patterns and loss mechanisms. In this way our
microtube ring resonators open a wide field of fundamental research
and practical applications.\\ \indent We would like to thank Markus
Br\"{o}ll, Stephan Schwaiger and Yuliya Stark for help during the
tube preparation and Jan Podbielski and Dirk Grundler for fruitful
discussions. We acknowledge financial support by the DFG via Grant
Nos. SFB~668, SFB~508, and GrK~1286.

\newpage

FIG 1 (color online) (a) Diagram of the strained
$\textrm{Ni}_{80}\textrm{Fe}_{20}/\textrm{GaAs}/\textrm{In}_{20}\textrm{Ga}_{80}\textrm{As}$
layer system used to prepare the RUPTs. The strained layer system is
released from the substrate by selectively etching the sacrificial
AlAs layer. (b) Sketch of a RUPT. It is positioned on the signal
line (S) between the two ground lines (G) of a coplanar waveguide.
Note the spiral geometry of the RUPT. (c) SEM image of RUPT A on the
signal line of a coplanar waveguide. A window was prepared into the
RUPT after the measurements were performed. This allowed an analysis
of the RUPTs internal structure (see inset). The RUPT is formed of
$3.5$ tight windings of the strained layer system.\\

FIG 2 (a) Microwave power absorption spectra for RUPT~A at $H=0$.
Four successive resonances can be identified. (b)~Resonance
frequency for the three larger peaks (squares, circles and
triangles). The fourth peak can only be identified for $H={0}$, but
is indistinguishable for higher external magnetic fields (diamonds).
The modeled dispersion for interfering spin waves, each fulfilling
the condition $k_{\phi}=2n/d$ is plotted for $n=0$ (continuous
line), $n=1$ (dashed), $n=2$ (dotted) and $n=3$ (dashed and dotted)
as described in the text. (c) Microwave power absorption spectra for
RUPT B at $H=0$, with four resonance peaks as well. (d) Resonance
frequency for the three larger peaks (squares, circles and
triangles) are shown up to $H=60$~mT, while the fourth peak can only
be resolved up to $H=25$~mT. The dispersion was fitted with the same
model as used for Fig. 2(b).\\

FIG 3 (color online) Calculated magnetic dispersion for RUPT A at
$H=0$ with an effective thickness~$t_{\textrm{eff}} = 37$~nm and an
effective demagnetizing field of $H_{\textrm{dem}} = 16$~mT. The
dotted lines mark the wavevector for modes $n=0,1,2,3...$ and the
corresponding resonant frequencies. The inset color plots show the
phase distribution in the RUPT for $n=0,1,2,3...$ (red and blue
represent opposite phase).\\

FIG 4 (a) Magnetic dispersion relation measured with broadband
microwave spectroscopy for external fields $H$ applied
perpendicularly to the axis of RUPT A. (b) Corresponding measurement
for RUPT B. In both cases, the resonance frequency reaches a
minimum, from which we can extract the absolute value of the
effective demagnetizing field $H_{\textrm{dem}}$.\\

\newpage

\begin{figure}
\begin{center}
\includegraphics[width=12cm]{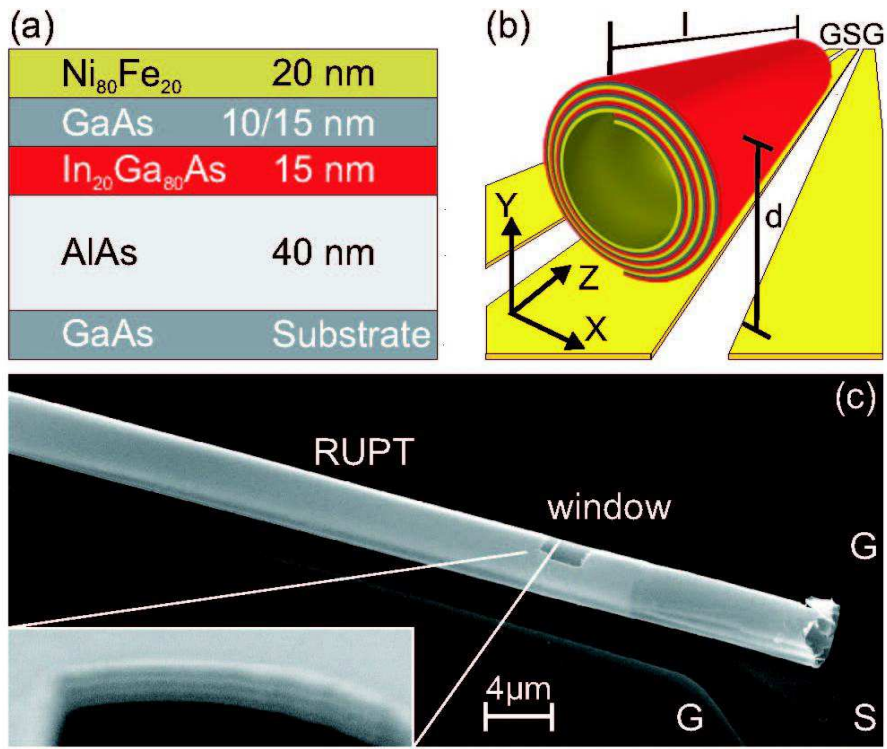}
\caption{(color online)} \label{figure1}
\end{center}
\end{figure}
\newpage
\begin{figure}
\begin{center}
\includegraphics[width=12cm]{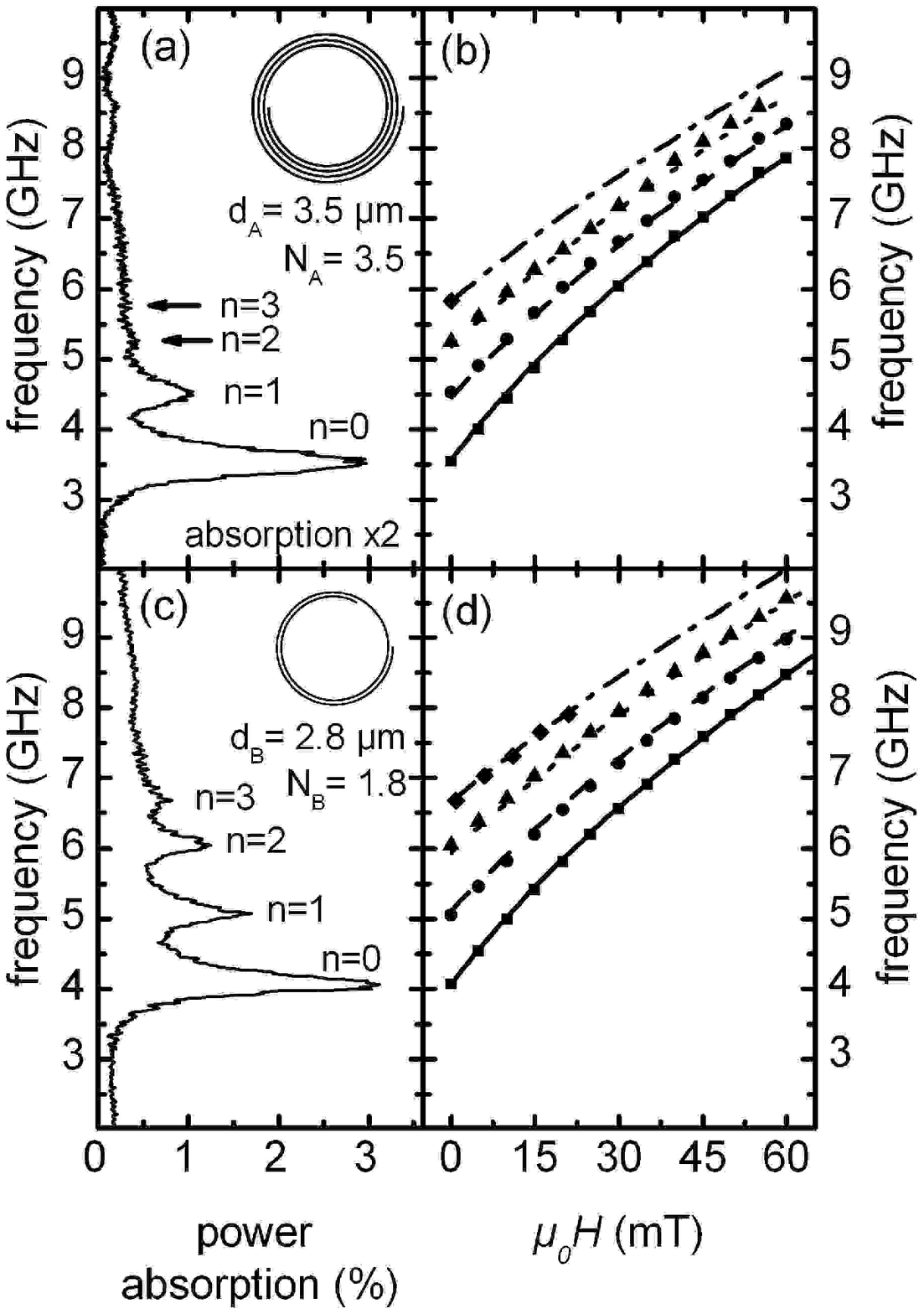}
\caption{} \label{figure2}
\end{center}
\end{figure}
\newpage
\begin{figure}
\begin{center}
\includegraphics[width=12cm]{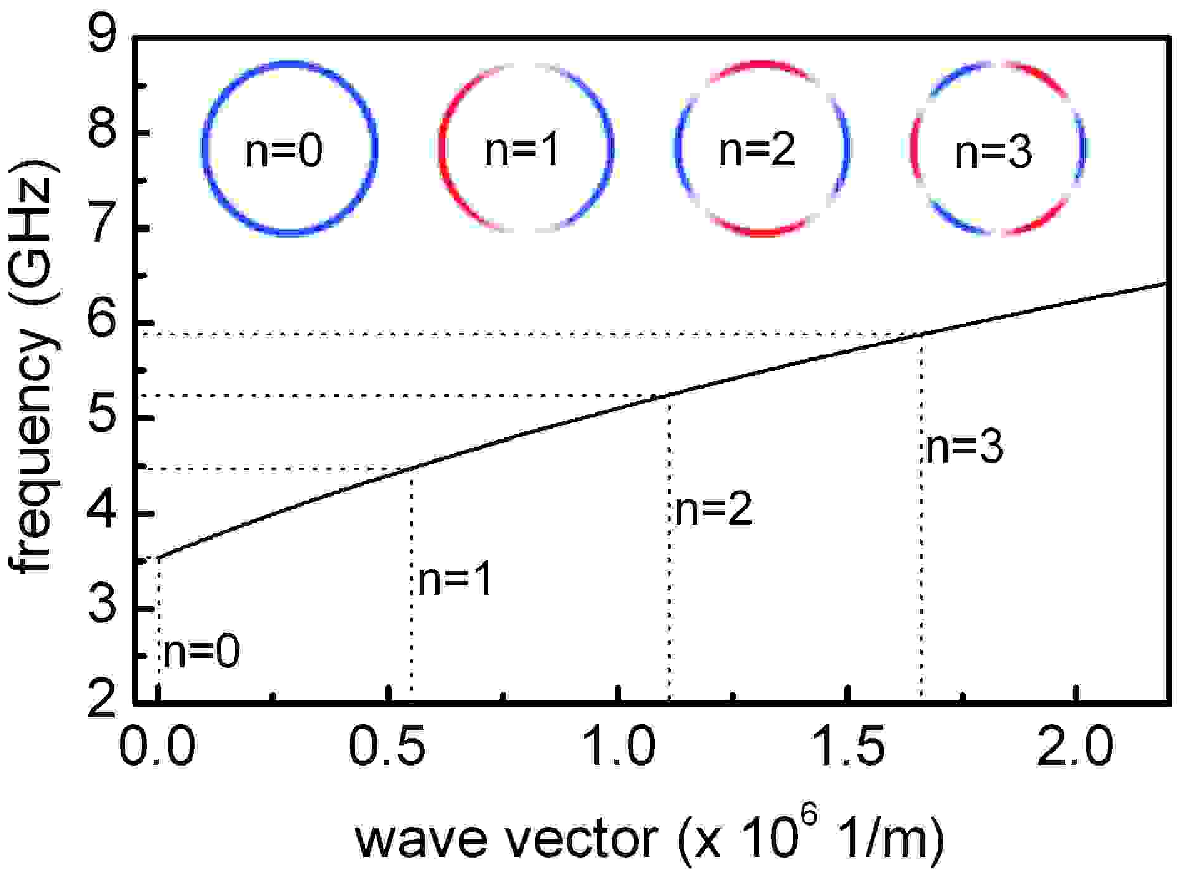}
\caption{(color online)} \label{figure3}
\end{center}
\end{figure}
\newpage
\begin{figure}
\begin{center}
\includegraphics[width=12cm]{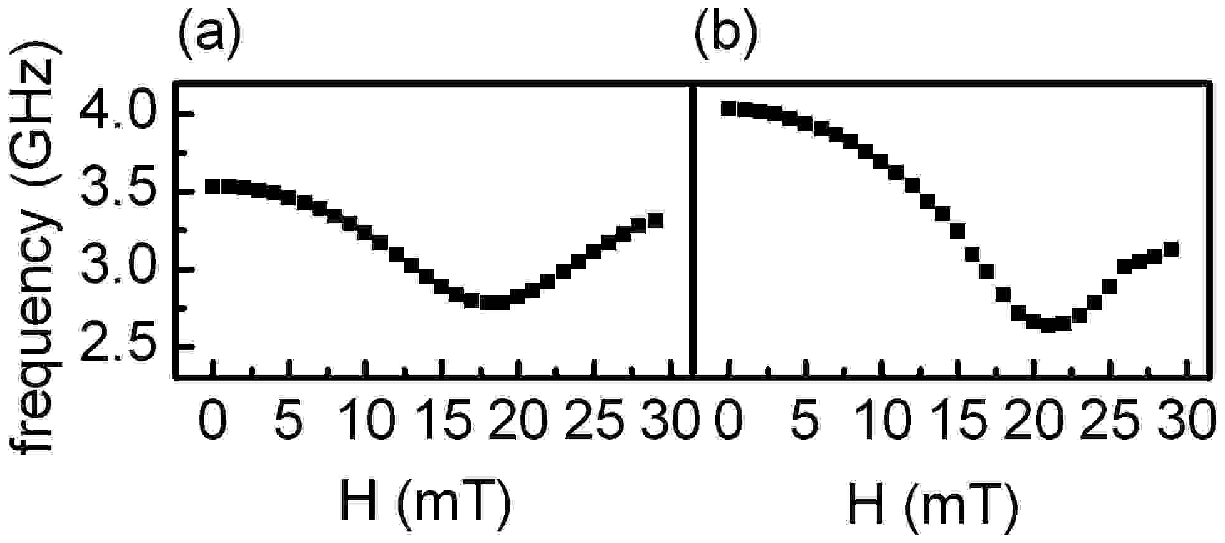}
\caption{} \label{figure4}
\end{center}
\end{figure}

\end{document}